\documentclass[aps,12pt,preprintnumbers,nofootinbib,superscriptaddress]{revtex4}
\usepackage{graphicx}
\usepackage{amssymb,amsmath}
\def\lag{{\mathcal{L}}}

\begin{document}
\newcount\hour \newcount\minute
\hour=\time \divide \hour by 60
\minute=\time
\count99=\hour \multiply \count99 by -60 \advance \minute by \count99
\newcommand{\mydate}{\ \today \ - \number\hour :00}

\preprint{CALT 68-2720}\preprint{UCSD PTH 09-02}

\title{Lee-Wick Theories at High Temperature}

\author{ Bartosz Fornal}
\email[]{ fornal@theory.caltech.edu}
\affiliation{California Institute of Technology, Pasadena, CA 91125}

\author{ Benjam\'\i{}n Grinstein}
\email[]{ bgrinstein@ucsd.edu}
\affiliation{Department of Physics, University of California at San Diego, La Jolla, CA 92093}

\author{ Mark B. Wise}
\email[]{ wise@theory.caltech.edu}
\affiliation{California Institute of Technology, Pasadena, CA 91125}

\date{\today}

\begin{abstract}

An extension of the standard model, the Lee-Wick standard model,
based on ideas of Lee and Wick was recently introduced. It does not
contain quadratic divergences in the Higgs mass and hence solves the
hierarchy puzzle. The Lee-Wick standard model contains new heavy
Lee-Wick resonances at the TeV scale that decay to ordinary
particles. In this paper we examine the behavior of Lee-Wick
resonances at high temperature. We argue that they contribute
negatively to the energy density $\rho$ and pressure $p$ and at
temperatures much greater than their mass $M$ their ${\cal O}(T^4)$
contributions to $\rho $ and $p$ cancel against those of the
ordinary (light) particles. The remaining ${\cal O}(M^2T^2)$
contributions are positive and result in an equation of state that
approaches $w=1$ from below as $T\rightarrow \infty$.
\end{abstract}

\maketitle
\newpage


\section{Introduction}

Recently ideas proposed by Lee and Wick~\cite{Lee:1969fy,Lee:1970iw}
were used to extend the standard model so that it does not contain
quadratic divergences in the Higgs mass
\cite{Grinstein:2007mp}. Higher derivative kinetic terms are added for
each of the standard model fields. They improve the convergence of
Feynman diagrams and result in a theory where there are no
quadratically divergent radiative corrections to the Higgs mass. The
higher derivative terms give rise to propagators with new poles that
are massive resonances. These Lee-Wick (LW) resonances have wrong-sign
kinetic terms which naively give rise to unacceptable instabilities
and violations of unitarity. Lee and Wick~\cite{Lee:1969fy,Lee:1970iw}
and Cutkowski {\it et al.} (CLOP) \cite{Cutkosky:1969fq} proposed a
way of defining the integrations that arise in Feynman diagrams so
that the theory is unitary, Lorentz invariant, and free of
instabilities. However, there is acausal behavior caused by the
unusual location of poles in the propagators. Physically this
acausality is associated with the future boundary condition needed to
forbid the exponentially growing modes. As long as the masses and
widths of the LW resonances are large enough, this acausality does not
manifest itself on macroscopic scales and is not in conflict with
scattering experiments. Various aspects of this
model \cite{Espinosa:2007ny,Dulaney:2007dx}, its
extensions \cite{Antoniadis:2007xc,Carone:2008bs} and of Lee-Wick
theories in
general \cite{Grinstein:2007iz,Grinstein:2008qq,Grinstein:2008bg,vanTonder:2008ub,Carone:2008iw}
have been explored in the recent literature. Collider
phenomenology \cite{Rizzo:2007ae,Krauss:2007bz,Rizzo:2007nf},
constraints from electroweak precision
measurement \cite{Alvarez:2008za,Underwood:2008cr,Carone:2008bs} and
the cosmology of theories with higher derivatives \cite{Cai:2008qw} have also been
studied.

In this paper we examine the high temperature behavior of Lee-Wick
theories, including the LW standard model. In these theories the
$S$-matrix can be calculated in perturbation theory using the
prescriptions of Lee and Wick and CLOP. It is unclear whether a functional integral formulation of LW theory exists, so a computation of finite temperature effects solely based on the known S matrix is desired. The formalism of Dashen, Ma
and Bernstein (DMB) \cite{Dashen:1969ep} expresses the thermodynamic grand
potential in terms of the $S$-matrix and we apply it to LW theories to
deduce the pressure and energy density for these theories at finite
temperature. Although previous analyses have argued that in scattering
experiments no acausal effects persist to macroscopic scales, it is
interesting to examine whether this is possible when multiple
scattering effects play a role. This is the case for thermal
equilibrium and we explore the propagation of sound waves in a gas
consisting of ordinary and Lee-Wick particles. We find that in such a
gas, at a large (but finite) temperature, sound waves propagate at a
speed less than light.

In the next section we review scattering in a simple scalar Lee-Wick
theory.  Section 3 uses the DMB formalism to calculate, in this toy
model, the energy density and pressure at thermal equilibrium. At high
temperatures $T\gg M$ we find that the LW resonance contributes minus
what an ordinary particle of mass $M$ would.  We use our results to
conclude that as $T \rightarrow \infty$ the speed of sound approaches
$c_s=1$ from below.  This gives in the limit $T \rightarrow \infty$ a speed of sound equal to the speed of
light and is the largest value consistent with causal
propagation of classical sound waves in the gas. Concluding remarks
are made in Section 4.

\section{A Toy Model}

In this section we introduce a simple Lee-Wick theory with a single
self-interacting real scalar field. In addition to the standard
kinetic term there is a higher derivative term. The Lagrangian density
is
\begin{equation}
\lag = \frac{1}{2} \partial_\mu \hat \phi \partial^\mu \hat \phi - \frac{1}{2 M^2} (\partial^2 \hat \phi)^2 - \frac{1}{2} m^2 \hat \phi^2 - \frac{1}{3!} g \hat \phi^3 ,
\end{equation}
so the propagator of $\hat \phi$ in momentum space is given by
\begin{equation}
D_F(p) = \frac{i}{p^2 - p^4/M^2 - m^2} .
\end{equation}
For $M \gg m$, this propagator has poles at $p^2 = m^2$ and also at $p^2 =
M^2$. Thus, the propagator describes more than one degree of freedom.

We can make these new degrees of freedom manifest in the Lagrangian
density in a simple way. First, let us introduce an auxiliary scalar field
$\tilde \phi$, so that we can write the theory as
\begin{equation}
\lag =  \frac{1}{2} \partial_\mu \hat \phi \partial^\mu \hat \phi  - \frac{1}{2} m^2 \hat \phi^2 - \tilde \phi \partial^2 \hat \phi + \frac{1}{2}M^2 \tilde \phi^2 - \frac{1}{3!} g \hat \phi^3 .
\end{equation}
Next, we define $\phi = \hat \phi + \tilde \phi$. In terms of this
variable, after integrating by parts, the Lagrangian density becomes
\begin{equation}
\lag = \frac{1}{2} \partial_\mu \phi \partial^\mu \phi - \frac{1}{2} \partial_\mu \tilde \phi \partial^\mu \tilde \phi + \frac{1}{2} M^2 \tilde \phi^2 - \frac{1}{2} m^2 ( \phi - \tilde \phi)^2 - \frac{1}{3!} g (\phi - \tilde \phi)^3.
\label{eq:scalarthy}
\end{equation}
In this form, it is clear that there are two kinds of scalar fields: a
normal scalar field $\phi$ and a new field $\tilde \phi$, which we
will refer to as a LW field. The sign of the quadratic Lagrangian of
the LW field is opposite to the usual sign so one may worry about
stability of the theory, even at the classical level. We will return
to this point. If we ignore, for simplicity, the mass $m$, the
propagator of $\tilde \phi$ is given by
\begin{equation}
\tilde D_F(p) = \frac{-i}{p^2 - M^2} .
\end{equation}
The LW field is associated with a non-positive definite norm
on the Hilbert space, as indicated by the unusual sign of its
propagator. Consequently, if this state were to be stable, unitarity of
the $S$-matrix would be violated. However, as emphasized by Lee and Wick,
unitarity and Lorentz invariance can be preserved provided that $\tilde \phi$ may decay. This is
natural in the theory described by Eq. (\ref{eq:scalarthy}) because $\tilde
\phi$ is heavy and can decay into two $\phi$ particles.

In the presence of the mass $m$, there is a mixing between the scalar
field $\phi$ and the LW scalar $\tilde \phi$. We can diagonalize this
mixing without spoiling the diagonal form of the derivative terms by
performing a hyperbolic rotation of the fields: $\phi=\phi^\prime
\cosh \theta + {\tilde \phi}^\prime \sinh \theta $, ${\tilde
  \phi}=\phi^\prime\sinh \theta + {\tilde \phi}^\prime\cosh \theta $.
This transformation diagonalizes the Lagrangian if
\begin{equation}
\tanh 2 \theta = \frac{-2 m^2/M^2}{1- 2 m^2/M^2}.
\end{equation}
A solution for the angle $\theta$ exists provided $M > 2m$. The
Lagrangian density~\eqref{eq:scalarthy} describing the system becomes
\begin{equation}
\lag = \frac{1}{2} \partial_\mu \phi^\prime \partial^\mu \phi^\prime - \frac{1}{2} m^{\prime2} \phi^{\prime2}
- \frac{1}{2} \partial_\mu \tilde \phi^\prime \partial^\mu \tilde \phi^\prime
+ \frac{1}{2} M^{\prime2} \tilde \phi^{\prime2} - \frac{1}{3!}g' (\phi^\prime - \tilde \phi^\prime)^3 ,
\end{equation}
where $m^\prime$ and $M^\prime$ are the masses of the diagonalized
fields and $g^\prime = (\cosh \theta - \sinh \theta)^3 g$. In what
follows we assume that $M \gg m$, so that $g^\prime \simeq
g$.

Introducing the LW fields makes the physics of the theory clear. There
are two fields; the heavy LW scalar decays to lighter scalars. At
loop level, the presence of the heavier scalar improves the convergence
of loop graphs at high energy consistent with our expectations from the
higher derivative form of the theory.

Loop corrections to the two point function of the LW
field play a crucial role. Near $ p^2 =M^2$ and at small $g$ the
${\tilde \phi}-\phi$ mixing can be neglected and the full LW $\tilde
\phi$ propagator and its perturbative expansion are given by
\begin{align}
\tilde D_F(p) &=\frac{-i}{p^2 - M^2} + \frac{-i}{p^2 - M^2} \left[i \Sigma(p^2)\right] \frac{-i}{p^2 - M^2}+\cdots \nonumber \\
&= \frac{-i}{p^2 - M^2 - \Sigma(p^2)} .
\end{align}
Note that, unlike for ordinary scalars, there is a minus sign in front
of the self-energy $\Sigma(p^2)$ in the denominator.  The pole mass
shift of the LW scalar coming from the radiative corrections is
$+\Sigma(M^2)$.  This sign is significant; for example, from a
one-loop computation we see that the imaginary part of the self energy
is
\begin{equation}
{\rm Im}\Sigma(p^2)={g^2 \over {32 \pi} }\theta(p^2- 4 m^2)\sqrt{1- \frac{4 m^2}{p^2}}.
\end{equation}
Therefore the propagator develops a pole for ${\rm Im}(p^2)>0$. In the narrow width approximation the propagator for the LW field is
\begin{equation}
D_{\rm LW}={ -i \over p^2-M^2 +iM\Gamma},
\end{equation}
where
\begin{equation}
\label{LWG}
\Gamma = -\frac{g^2}{32 \pi M} \sqrt{1- \frac{4 m^2}{M^2}}.
\end{equation}
This width differs in sign from widths of the usual unstable particles
we encounter. Strictly speaking the propagator has an additional pole
at $p^2=M^2+iM\Gamma$ and a cut over the real axis; however, the effect
of these two terms in the below calculation of the pressure and energy
density at finite temperature cancel one another so we ignore them.

Using the Lee and Wick and CLOP prescriptions the $S$-matrix in this
theory is unitary and Lorentz invariant on the space of physical
ordinary $\phi$ particles. The contour of integration over $p^0$ in this prescription does not lead to any instability, as one may naively guess from  the negative sign in the width, but instead leads to apparently
acausal behavior. This has been extensively discussed in the
literature \cite{Lee:1969fy,Grinstein:2008bg,Coleman:1969xz}.

The LW resonance $\tilde \phi$ is unstable and therefore does not
appear in the initial or final states of the $S$-matrix. This is
similar to the case of the $W$-boson of the standard model, which does
not appear in initial or final states of $S$-matrix elements because
it is unstable\footnote{One difference is that there are no
  poles in amplitudes associated with the $W$-boson two point
  function, just a cut that is represented as a pole in the narrow
  width approximation. However, for the Lee-Wick $\tilde \phi$
  resonance there is actually a pair of negative residue poles in the complex plane in
  addition to the usual cut.}. Nonetheless, the $\tilde \phi$
resonance impacts the scattering of ordinary $\phi$ particles
$\phi(p_1)+\phi(p_2) \rightarrow \phi(p_1')+\phi(p_2')$, particularly
near the kinematic point $(p_1+p_2)^2=M^2$.

Writing $S=1-i{\cal T}$, the ${\cal T}$ matrix can be computed using the standard
Feynman techniques, modified appropriately for Lee-Wick theories. DMB
introduce a closely related quantity ${\cal T} (E)$ that has two
particle matrix elements
\begin{equation}\label{T}
\langle {\tilde {\bf p}}_1,{\tilde {\bf p}}_2|{\cal T}(E)| {\bf p}_1,{\bf p}_2 \rangle = (2 \pi) \delta(E-E_1-E_2) (2 \pi)^3\delta^3({\bf P}-{\tilde {\bf P}}) {\cal M}(E),
\end{equation}
where ${\bf P}={\bf p}_1+{\bf p}_2$, ${\tilde {\bf P}}={\tilde {\bf
    p}}_1+{\tilde{\bf p}}_2$ and $\cal M$ is essentially the usual
invariant matrix element. For center of mass energies near $M$ the
invariant matrix element is given (in the narrow resonance
approximation) by\footnote{There is an extra factor of 1/2 associated
  with identical particles that we have chosen to put in $\cal M$
  rather than in phase space integrations.}
\begin{equation}
{\cal M}(E)=-{1 \over 2}{g^2 \over E^2-{\bf P}^2 -M^2+iM{\Gamma}}.
\end{equation}
Note that this differs from scattering via the exchange of an ordinary ({\it i.e.}, not LW) resonance by an overall minus sign and the fact that $\Gamma$ given by Eq.~(\ref{LWG}) is negative.

\section{The Pressure and Energy Density in Thermal Equilibrium}

The grand partition function $\Omega$ at zero chemical potential is defined by
\begin{equation}
\Omega ={1 \over \beta} {\rm ln}{\rm Tr} e^{-\beta H},
\end{equation}
where $\beta=1/kT$ and the trace is over all physical states in the
theory. In our toy model the physical states are the $\phi$ particle
states but not states that contain a LW resonance. From $\Omega$ one
can calculate the thermal equilibrium pressure $p$ and energy density
${\rho}$ in the usual fashion using formulas
\begin{equation}
p=-{\Omega \over V}\ ,~~~~~ \rho= -{\partial(\beta p) \over \partial \beta},
\end{equation}
where $V$ is the volume of the system.

DMB derive the following expression for the grand potential
\begin{equation}
\label{grandpotential}
\Omega=\Omega_0-{1 \over \beta} \int {\rm d}E e^{-\beta E} {1 \over 4 \pi i}\left( {\rm Tr} A S(E)^{-1}\overleftrightarrow{\frac{\partial}{\partial{E}}} S(E)\right)_c,
\end{equation}
where $c$ denotes that only connected diagrams are taken into account.
In Eq.~(\ref{grandpotential}) the $S$-matrix is given by
$S(E)=1-i{\cal T}(E)$, $\Omega_0$ is the free particle grand potential
and $A$ is an operator that sums over permutations of the identical
particles in the trace with the appropriate minus signs for
fermions. Using the relation between $S$ and $\cal T$ this becomes
\begin{equation}
\label{DMB}
\Omega=\Omega_0-{1 \over \beta} \int {\rm d}E e^{-\beta E} {1 \over 4 \pi i}\left[{\rm Tr}A\left(-i{\partial \over \partial E} \left[{ \cal T}(E)+{ \cal T}(E)^{\dagger}\right]+  {\cal T}(E)^{\dagger}\overleftrightarrow{\frac{\partial}{\partial{E}}} {\cal T}(E)\right)\right]_c.
\end{equation}
To evaluate $\Omega$ in the toy model introduced in the previous
section, we begin by evaluating the part of $\Omega$ that comes from
the contribution of two particle $\phi$ states. It is convenient to
use the phase space relation
\begin{equation}
\label{phasespace}
\int {{\rm d}^3 p_1 \over (2 \pi)^3 2 E_1}\int{{\rm d}^3  p_2 \over (2 \pi)^3 2 E_2}=\int{\rm d}^3 P \int {{\rm d}^3 p_1' \over (2 \pi)^3 2E_1'}
\int{{\rm d}^3 p_2' \over (2 \pi)^3 2 E_2'}\delta^3({\bf p'}_1+{\bf p'}_2) {\omega \over E},
\end{equation}
where the primed variables are the center of mass momenta and
energies, $\omega= E_1'+E_2'$ and $E=E_1+E_2=\sqrt{\omega^2+{\bf
    P}^2}$. In order to calculate the first term of the integral in
Eq.~(\ref{DMB}) we notice that\footnote{For simplicity we drop the
  subscript $c$.}
\begin{equation}
{\rm Tr }{\partial \over \partial E}{\cal T}(E)={\partial \over \partial E}{\rm Tr }{\cal T}(E),
\end{equation}
so the expression we need to evaluate is
\begin{equation}
{\rm Tr }{\cal T}(E)=\int {{\rm d}^3 p_1 \over (2 \pi)^3 2 E_1}\int{{\rm d}^3  p_2 \over (2 \pi)^3 2 E_2} (2 \pi)^3\delta^3(0)(2 \pi)\delta(E-E_1-E_2){\cal M}(E).
\end{equation}
Using $(2 \pi)^3\delta^3(0)=V$ and the phase space relation (\ref{phasespace}) gives
\begin{equation}
{\rm Tr }{\cal T}(E)=V\int \frac{{\rm d}^3P}{(2\pi)^3}{\cal M}(E)\int  {{\rm d}^3 p_1' \over (2 \pi)^3 2E_1'}\int {{\rm d}^3 p_2' \over (2\pi)^3 2E_2'} \delta^3({\bf p'}_1+{\bf p'}_2) {\omega \over E}(2 \pi)^4 \delta(E-E_1-E_2).
\end{equation}
Now, recall that $P^{\mu}=(E,{\bf P})$ is the total energy-momentum
four-vector of the states in the trace while $(\omega,{\bf 0})$ is the
corresponding energy-momentum four-vector in the center of mass
frame. They are related by a boost with a Lorentz gamma factor
$\gamma=E/{\omega}$. One can go from one set of variables to the
other. Using the relation $E^2=\omega^2+{\bf P}^2$ we have that
\begin{equation}
{\omega \over E} \delta(E-E_1-E_2)=\delta(\omega-E_1'-E_2')
\end{equation}
and so the integrations over ${\rm d}^3 p_1'{\rm d}^3 p_2'$ become the
standard two body phase space integration. We arrive at the result
\begin{equation}
{\rm Tr }{\partial \over \partial E}{\cal T}(E)=V{\partial \over \partial E}\int{{\rm d}^3 P \over (2\pi)^3}{1 \over 8 \pi}\sqrt{1-{4 m^2 \over \omega^2}}{\cal M}(E).
\end{equation}
Finally, we change from the variable $E$ to $\omega$ in all other
places. Using
\begin{equation}
{\rm d}E {\partial \over \partial E}={\rm d}\omega {\partial \over \partial \omega}
\end{equation}
and  interchanging the order of the two integrations we get
\begin{equation}
\int {\rm d}E e^{-\beta E}{\rm Tr }{\partial \over \partial E}{\cal T}(E)=\int {{\rm d}^3 P \over (2\pi)^3}\int {\rm d}\omega e^{-\beta\sqrt{\omega^2 +{\bf P}^2}}{\partial \over \partial \omega}\left({1 \over 8 \pi}\sqrt{1-{4m^2\over \omega^2}}{\cal M}(\omega)\right),
\end{equation}
where
\begin{equation}
{\cal M}(\omega)=-{1 \over 2}{g^2 \over \omega^2 -M^2+iM{\Gamma}}.
\end{equation}
The term with ${ \cal T}(E)^{\dagger}$ gives the same contribution to
$\Omega$ but with ${\cal M}(\omega)$ substituted by ${\cal
  M}^*(\omega)$.  In calculating the second term of the integral in
Eq.~(\ref{DMB}) we use the same methods as previously.  We need to
evaluate
\begin{eqnarray}
&&{\rm Tr}{\cal T}(E)^{\dagger}\overleftrightarrow{\frac{\partial}{\partial{E}}} {\cal T}(E)
=V\int \prod_{i=1}^2 {{\rm d}^3 p_i \over (2 \pi)^3 2 E_i}
\int \prod_{i=1}^2 {{\rm d}^3 \hat{p}_i \over (2 \pi)^3 2 \hat{E}_i}(2 \pi)^3\delta^3({\bf P}-{\hat {\bf P}})\nonumber \\
&&\left[(2 \pi)\delta(E-E_1-E_2){\cal M^*}(E)\right]
\overleftrightarrow{\frac{\partial}{\partial{E}}}\left[(2 \pi)\delta(E-\hat{E}_1-\hat{E}_2){\cal M}(E)\right],
\end{eqnarray}
where ${\bf P}={\bf p}_1+{\bf p}_2$, ${\hat {\bf P}}={\hat {\bf
    p}}_1+{\hat{\bf p}}_2$ and the factor $V(2 \pi)^3\delta^3({\bf
  P}-{\hat {\bf P}})$ came from the momentum delta functions in the
definition (\ref{T}) of ${\cal T}(E)$.  In the c.m. frame we have
\begin{eqnarray}
&&{\rm Tr}{\cal T}(E)^{\dagger}\overleftrightarrow{\frac{\partial}{\partial{E}}} {\cal T}(E)
=V\int {{\rm d}^3 P \over (2 \pi)^3}\int {{\rm d}^3 \hat{P} \over (2 \pi)^3}(2 \pi)^3\delta^3({\bf P}-{\hat {\bf P}})
\left[\int \prod_{i=1}^2 {{\rm d}^3 p_i' \over (2 \pi)^3 2 E_i'}\delta^3({\bf p'}_1+{\bf p'}_2) {\omega \over E} \right.\nonumber \\
&&\left.(2 \pi)^4\delta(E-E_1-E_2){\cal M^*}(E)\right]
\overleftrightarrow{\frac{\partial}{\partial{E}}}\left[\int \prod_{i=1}^2 {{\rm d}^3 \hat{p}_i' \over (2 \pi)^3 2 \hat{E}_i'}
\delta^3(\hat{{\bf p}}'_1+\hat{{\bf p}}'_2) {\hat{\omega} \over E}
(2 \pi)^4\delta(E-\hat{E}_1-\hat{E}_2){\cal M}(E)\right]=\nonumber\\
&&=V\int {{\rm d}^3 P \over (2 \pi)^3}\left[\frac{1}{8\pi}\sqrt{1-\frac{4m^2}{\omega^2}}{\cal M^*}(E)\right]\overleftrightarrow{\frac{\partial}{\partial{E}}}\left[\frac{1}{8\pi}\sqrt{1-\frac{4m^2}{\omega^2}}{\cal M}(E)\right].
\end{eqnarray}
Putting all this together and neglecting the mass $m$ of the ordinary
scalars, we arrive at the LW contribution to the grand potential in
the form
\begin{equation}
\Omega_{\rm LW}=-{V \over \beta}\int {{\rm d}^3 P\over (2\pi)^3}\int {\rm d}\omega e^{-\beta\sqrt{\omega^2 +{\bf P}^2}}{1 \over 4 \pi i} \left[-i{\partial \over \partial \omega} \left({{\cal M}(\omega) \over 8 \pi}+{{\cal M}^*(\omega) \over 8 \pi}\right) +{{\cal M}^*(\omega) \over 8 \pi}\overleftrightarrow{\frac{\partial}{\partial{\omega}}}{{\cal M}(\omega) \over 8 \pi}\right]+\ldots,
\end{equation}
where the ellipses denote the terms from summing over permutations
which basically are multiple insertions of the two body
state. Performing the differentiations and using the explicit formulas
for ${\cal M}$ and $\Gamma$ this becomes
\begin{equation}
\Omega_{\rm LW}=-{V \over \beta}\int {{\rm d}^3 P\over (2\pi)^3}\int {\rm d}\omega e^{-\beta\sqrt{\omega^2 +{\bf P}^2}} \left[{2 \over \pi}{\omega M \Gamma \over (\omega^2-M^2)^2 +M^2 \Gamma^2}\right]+\ldots.
\end{equation}
Recall that for the LW resonance $\Gamma$ is negative. The above
formula is the same as one would get for scattering through an
ordinary resonance except in that case $\Gamma$ is
positive. Therefore, in the narrow LW resonance approximation
\begin{equation}
{2\over \pi} {\omega M \Gamma \over (\omega^2-M^2)^2 +M^2 \Gamma^2}\rightarrow -\delta(\omega-M).
\end{equation}
Hence, the contribution to the grand potential from the single LW resonance is
\begin{equation}
\Omega_{\rm single\ LW}=+{V \over \beta}\int {{\rm d}^3 P\over (2\pi)^3}e^{-\beta\sqrt{M^2+{\bf P}^2}}.
\end{equation}
This is precisely what one would expect from a stable particle of mass
$M$, except for the overall plus sign instead of a minus. Note that the narrow width
approximation is valid provided  the prefactor in the integral over
$\omega$ is slowly varying. This will be the case provided
$\beta|\Gamma|\ll1$.

Thus far we have only included two particle states in the calculation of the trace for the grand potential, 
with a resonant S matrix from  Eq.~(13). Contributions to the trace from states with more than two particle 
introduce multiple resonance amplitudes. This has two effects \cite{Dashen:1974jw}:
they modify the width, as expected for decays in a thermal bath, and
they convert the exponential factor to the usual Bose-Einstein
logarithm. In the narrow width approximation the modified width is
still narrow, that is, proportional to $g^2$, and hence it still
vanishes as $g\to0$. The Bose-Einstein logarithm arises from
considering multiple resonance graphs that are connected only because
of the permutation operator $A$. This gives
\begin{equation}
\Omega_{\rm LW}=-{V \over \beta}\int {{\rm d}^3 P\over (2\pi)^3}{\rm ln}\left(1-e^{-\beta\sqrt{M^2+{\bf P}^2}}\right),
\end{equation}
which is minus the contribution of a boson of mass $M$ to the ideal gas grand potential.
Note that this result is valid for arbitrarily large temperatures, which
is not the case for non-elementary (composite) resonances
\cite{Dashen:1974yy}.

Since in the narrow width approximation the LW resonance contributes minus
what an ordinary scalar particle of mass $M$ would, in our toy
model the LW contribution to the energy density is
\begin{equation}
\rho_{\rm LW}=-\left[{\pi^2 (kT)^4 \over 30}-{M^2 (k T)^2 \over 24 }\right] +\ldots,
\end{equation}
while the contribution to the pressure is
\begin{equation}
p_{\rm LW}=-\left[{\pi^2 (kT)^4 \over 90}-{M^2 (k T)^2 \over 24} + {M^3 (kT) \over 12 \pi} \right] +\ldots.
\end{equation}
Here the ellipses stand for terms of order ${\rm ln}(T)$ at most and are less important than those
explicitly displayed when $T\gg M$. Adding these to the positive
contributions from the ordinary scalar (whose mass $m$ we neglect)
\begin{equation}
\rho_{\rm ordinary}={\pi^2 (kT)^4 \over 30}\ ,~~~~~ p_{\rm ordinary}={\pi^2 (kT)^4 \over 90},
\end{equation}
gives
\begin{equation}
\rho=\rho_{\rm ordinary}+\rho_{\rm LW}={M^2 (k T)^2 \over 24 }+\ldots
\end{equation}
and
\begin{equation}
p=p_{\rm ordinary}+p_{\rm LW}={M^2 (k T)^2 \over 24 }-{M^3 (kT) \over 12 \pi}+\ldots .
\end{equation}
A similar analysis holds for theories with a left handed fermion and
its LW partner. This is most easily seen in the higher derivative
formulation of the theory. In the auxiliary field formulation one
usually introduces left and right handed LW fermions but one of these
is dependent on the other through the equations of motion. In that
case\footnote{We take into account a factor of $2s+1$ in the formula
  for the grand partition function, where $s$ is the spin.}
\begin{equation}
\rho_{\rm LW}^{\rm F}=-\left[{7\pi^2 (kT)^4 \over 120}-{M^2 (k T)^2 \over 24 }\right] +\ldots,
\end{equation}
and
\begin{equation}
p_{\rm LW}^{\rm F}=-\left[{7\pi^2 (kT)^4 \over 360}-{M^2 (k T)^2 \over 24 }\right] +\ldots,
\end{equation}
where, similarly to the boson case, the ellipses denote terms of order ${\rm ln}(T)$ at most.
The absence of a term linear in T  for the fermion pressure and density is as in the normal case.
Adding the LW fermion contribution to the ordinary fermion energy density and pressure
\begin{equation}
\rho_{\rm ordinary}^{\rm F}={7\pi^2 (kT)^4 \over 120}\ ,~~~~~ p_{\rm ordinary}^{\rm F}={7\pi^2 (kT)^4 \over 360},
\end{equation}
gives
\begin{equation}
\rho^{\rm F}=\rho_{\rm ordinary}^{\rm F}+\rho_{\rm LW}^{\rm F}={M^2 (k T)^2 \over 24 }+\ldots
\end{equation}
and
\begin{equation}
p^{\rm F}=p_{\rm ordinary}^{\rm F}+p_{\rm LW}^{\rm F}={M^2 (k T)^2 \over 24 }+\ldots .
\end{equation}
From the above formulas for the pressure and energy density one can
calculate the factor $w$ in the equation of state $p=w\rho$ to be
\begin{eqnarray}
&&w=1-\frac{2}{\pi}\frac{M}{kT}+{\cal O}\left[\frac{{\rm ln}(T)}{T^2}\right]\ \ \ {\rm for\ bosons},\nonumber \\
&&w=1+{\cal O}\left[\frac{{\rm ln}(T)}{T^2}\right] \ \ \ {\rm for\ fermions}.
\end{eqnarray}
In Figure 1 we plot $w=p/\rho$ for both cases as a function of $kT/M$.
\begin{figure}\label{fig1}
\begin{center}
\begin{tabular}{cc}
\includegraphics[width=.65\textwidth]{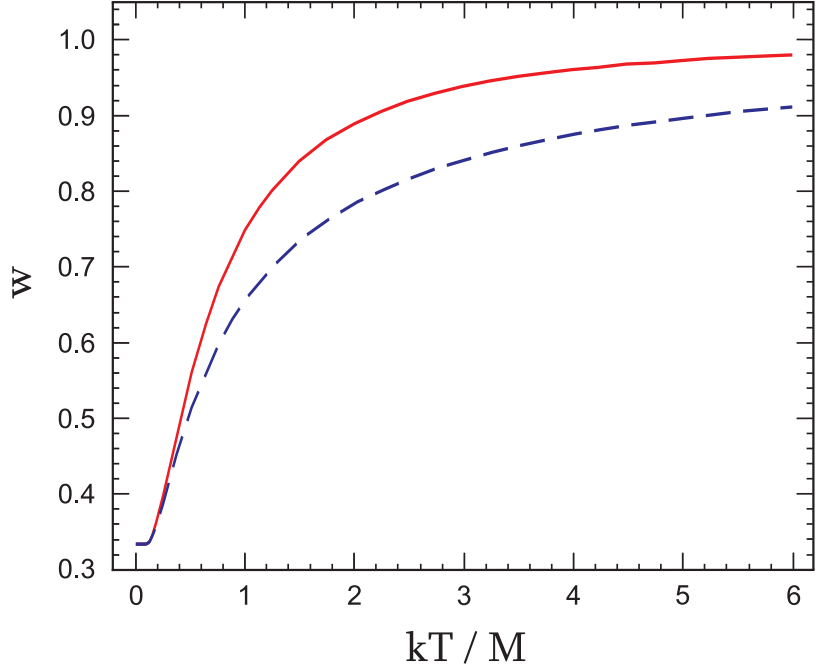}
\end{tabular}
\end{center}
\label{duration}
\caption{Factor $w$ in the equation of state $p=w\rho$ as a function of $kT/M$ for fermions (solid) and bosons (dashed).}
\end{figure}
The value $w=1/3$ for small temperatures should not be
puzzling since for $T\rightarrow 0$ the LW contribution is suppressed
by the Boltzmann weight factors and only the ordinary particles
contribute to the grand potential. More interesting is the value $w=1$
at high temperatures, which may have implications to the early
universe cosmology. Cosmology with equation of state $w=1$ has been
investigated in the context of holographic cosmology for a medium
consisting of a black hole emulsion \cite{Banks:2001px,Banks:2003ta,Banks:2004vg,Banks:2004cw,Banks:2004eb}.
\begin{figure}\label{fig2}
\begin{center}
\begin{tabular}{cc}
\includegraphics[width=.65\textwidth]{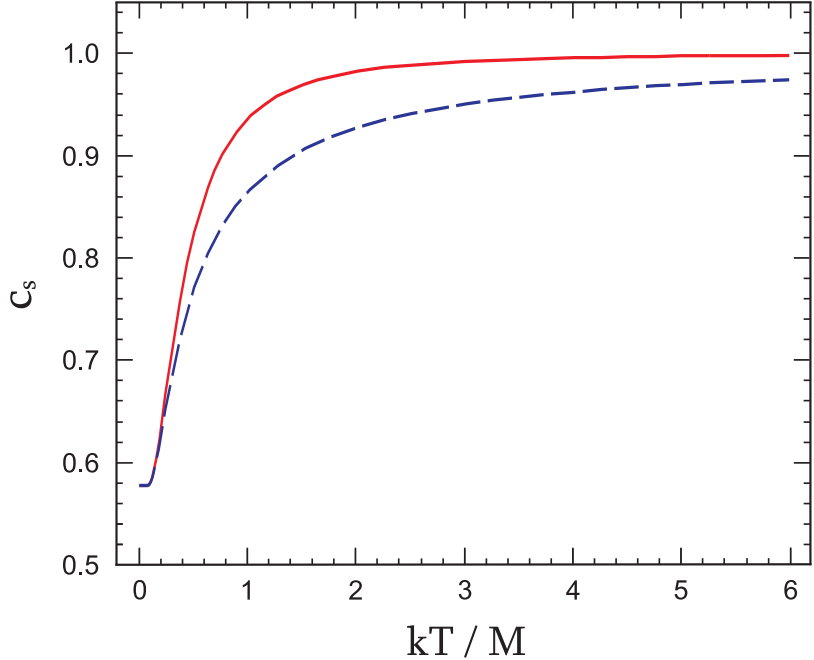}
\end{tabular}
\end{center}
\label{duration}
\caption{Speed of sound $c_s=\sqrt{{\rm d}p / {\rm d} \rho }$ as a function of $kT/M$ for fermions (solid) and bosons (dashed).}
\end{figure}

The speed of sound $c_s$ can be calculated using the formula
\begin{equation}
c_s=\sqrt{{\rm d}p \over {\rm d} \rho }= \sqrt{{{\rm d}p \over {\rm d}T } / {{\rm d} \rho \over {\rm d} T}} \ .
\end{equation}
Taking into account higher order correction terms we arrive at
\begin{eqnarray}
&&c_s=1-\frac{1}{2\pi}\frac{M}{kT}+{\cal O}\left[\frac{{\rm ln}(T)}{T^2}\right] \ \ \ {\rm for\ bosons},\nonumber \\
&&c_s=1+{\cal O}\left[\frac{{\rm ln}(T)}{T^2}\right] \ \ \ {\rm for\ fermions}.
\end{eqnarray}
Figure 2 shows the plot of $c_s$ for fermions and bosons as a function of $kT/M$.
The speed of sound increases from $1/\sqrt{3}$ at $T=0$ to $1$ as $T
\rightarrow \infty$, which is equal to the speed of light. Hence, the
propagation of classical sound waves in the relativistic gas is
causal.

In the LW standard model there is a LW partner for every ordinary
particle. Hence, at high temperatures above all the LW masses the
pressure and energy density are approximately proportional to $M^2
T^2$, which implies the same equation of state $w=1$. This gives a
universe with density $\rho \sim a(t)^{-6}$ and the scale factor $a(t)
\propto t^{1/3}$.

\section{Conclusions}

We have studied in this paper Lee-Wick theories at high
temperature. Making use of the $S$-matrix formulation of statistical
mechanics presented in DMB \cite{Dashen:1969ep}, we calculated the grand thermodynamical
potential for a gas of Lee-Wick resonances in the boson and fermion
case. We found that the contribution of Lee-Wick resonances to the
energy density and pressure is negative for high temperatures.

Next, we considered a gas of both ordinary and Lee-Wick particles. We
found that for high temperatures (much greater than the mass of the
resonance itself) the contributions of the Lee-Wick resonances to the
pressure and energy density cancel against those of the normal
particles at leading order in temperature ({\it i.e.}, ${\cal
  O}(T^4))$. We confirmed this for both fermions and bosons. We found
that the remaining ${\cal O}(M^2T^2)$ contribution is positive and
identical for the pressure and energy density. This led us to the
equation of state $w=1$ for $T\rightarrow\infty$. In applications to
big bang cosmology this  yields a
scale factor of the universe $a(t) \propto t^{1/3}$.

The quantity $c_s=\sqrt{{\rm d}p /{\rm d} \rho }$ corresponds to the speed of sound in the medium. We
checked that $c_s$ is less than 1 in the whole temperature range, thus
causality is not violated.  It seems interesting to investigate the
cosmological implications of the equation of state $w=1$, especially
for the propagation of fluctuations in the early universe.

\begin{acknowledgments}
  The work of BG and MBW was supported in part by the US Department of
  Energy under contracts DE-FG03-97ER40546 and DE-FG03-92ER40701,
  respectively. The work of BF was supported by the Henry and Grazyna
  A. Bauer Fellowship.
\end{acknowledgments}



\end{document}